\documentclass[11pt,a4paper]{article}\onecolumn
\usepackage[latin1]{inputenc} 
\usepackage{color}
\usepackage{graphicx,geometry}
\usepackage[footnotesize]{caption}
\usepackage{subcaption}
\usepackage{wrapfig}
\usepackage{txfonts}
\usepackage{booktabs}
\usepackage{pdflscape}
\usepackage{afterpage}
\usepackage{supertabular}
\usepackage{threeparttable}
\usepackage[affil-it]{authblk}

\title{Astronomy and Astrophysics in the Philosophy of Science\footnote{This is a draft of a chapter ``Astronomy and Astrophysics'' that has been accepted for publication by Oxford University Press in the forthcoming book ``The Oxford Handbook of Philosophy of Science'' edited by Paul Humphreys due for publication in May 2016.}}
\author{Sibylle Anderl$^{1, 2}$}
\affil{$^1$Univ. Grenoble Alpes, IPAG, F-38000 Grenoble, France\\$^2$CNRS, IPAG, F-38000 Grenoble, France\\sibylle.anderl@obs.ujf-grenoble.fr}
\date{}
  
\begin{document}
\maketitle

\abstract{This article looks at philosophical aspects and questions that modern astrophysical research gives rise to. Other than cosmology, astrophysics particularly deals with understanding phenomena and processes operating at ``intermediate'' cosmic scales, which has rarely aroused philosophical interest so far. Being confronted with the attribution of antirealism by Ian Hacking because of its observational nature, astrophysics is equipped with a characteristic methodology that can cope with the missing possibility of direct interaction with most objects of research. In its attempt to understand the causal history of singular phenomena it resembles the historical sciences, while the search for general causal relations with respect to classes of processes or objects can rely on the ``cosmic laboratory'': the multitude of different phenomena and environments, naturally provided by the universe. Furthermore, the epistemology of astrophysics is strongly based on the use of models and simulations and a complex treatment of large amounts of data.}
\vspace{0.5cm}



\section{Introduction}

Modern astrophysics\footnote{Often the transition from classical astronomy to today's astrophysics is denoted as the transition from purely descriptive astronomical observations to the explanatory application of physical methods to cosmic phenomena. However, in this article the terms ``astrophysics'' and ``astronomy'' will be used interchangeably. }, which operates at the ``intermediate'' scales (i.e. in between the physics of the solar system and the cosmology of the entire universe) has rarely been addressed directly within the philosophy of science context. While there are philosophical sub-disciplines in other special sciences, such as the philosophy of biology and the philosophy of chemistry, today's non-cosmological astrophysical research, until now, had failed to arouse philosophical interest in any lasting way. 

One reason for this lack of focused interest might have been that, when viewed from the outside, astrophysics superficially appears as a special case within the philosophy of physics or as an extension of the existing philosophy of cosmology. However, both fields seem to deal with slightly different core themes than the ones that come up in reflecting upon contemporary astrophysical research. In contrast with much of what gets discussed in the philosophy of physics, most astrophysical research programmes are not at all concerned with metaphysical questions. Rather than the development of autonomous mathematical theories that themselves then require specific interpretations, it is the application of existing physical theories to a vast array of cosmic phenomena and different environments that are found in the universe, which constitutes the main focus of astrophysical research. Understanding the non-reducible complexity that permeates the universe requires a broad knowledge of statistical samples rather than a detailed or specific understanding of the ontology of theoretical entities. Practical problems in astrophysical research are so deeply concerned with methodologies that arise from the observational constraints imposed on the empirical part of research, that ontological questions are seldom seen to be relevant. More than that, it might not even be clear what a notion of fundamentality could possibly mean with respect to objects of astrophysical research beyond the cases where astrophysics and microphysics meet and where foundational problems might be inherited from there. 

Cosmology, on the other hand, evokes its own metaphysical questions, e.g. when debates centre around the nature of space and time or the status of cosmological entities like dark matter and dark energy. Being the study of the origin and the evolution of the universe, cosmology has to face specific problems that arise from the totality of its research object: in studying the global properties of a single, uniquely defined object, possible cosmological knowledge is systematically underdetermined. This underdetermination is theoretical on the one hand, but also empirical on the other hand, because we cannot observe it in its entirety, even in principle. The reason is that light can only reach us from a limited region of the spacetime, given the limited speed of light and the finite age of the universe. However, to the extent that cosmology does not only lay claim to understanding the evolution of the universe as a whole, but also the evolution of its large-scale structures, the transition to non-cosmological astrophysical research that is concerned with the understanding of astrophysical phenomena on smaller scales is continuous. A philosophy of (non-cosmological) astrophysics and astronomy will therefore necessarily overlap topics from cosmological research. Some of the epistemological questions it faces might however be different from those that are classically discussed for the case of cosmology, resulting from a more bottom-up point of view as compared to the top-down case of a cosmology perspective.

Astrophysics and cosmology share a common problem in that they both need to acquire knowledge of their objects of research without directly interacting, manipulating or constraining them. In reconstructing plausible explanations and evolutionary scenarios of particular, observable objects and processes, the resulting methodology of astrophysics and cosmology resembles the criminology of Sherlock Holmes: the astrophysicist must look for all possible traces and clues that may help to illuminate what has happened in a given region of the universe. However, in contrast to the cosmological study of the present universe, astrophysics goes beyond the understanding of singular events and objects and aims at making general statements about classes of objects like O-type stars, the interstellar medium of galaxies, black holes or clusters of galaxies. Here, inability to stage real experiments imposes a challenge on the establishment of causal claims that may derive from but must transcend mere correlations. In practice, this can, at least partially, be compensated for by the fact that the universe is a ``Cosmic Laboratory'', that is filled with phenomena and on-going processes in all manner of evolutionary stages, each of them constrained by different initial conditions and contemporary environments. Accordingly, the laboratory scientist's skill in controlling her experiment has its correspondence in the astrophysicist's ability to statistically sample, analyze and model this diversity that she observes, in the complex process of reaching generalized conclusions about cosmic phenomena. The understanding of the epistemology of modern astrophysical research certainly can profit from epistemological studies of other scientific fields (such as paleontology, archaeology, and the social sciences) that also cannot easily perform experiments and artificially reduce their subject's complexity. At the same time, the ubiquitous use of simulations and models and the challenges that the ever increasing generation of the huge amount of data imposes on astrophysics make corresponding discussions from the philosophy of science relevant for the reflection on astrophysical research. 

Finally, astrophysical research is an interesting topic with respect to its practical organisation and realization, which opens the discussion to historical and sociological directions. The generation of  empirical astrophysical knowledge relies on the distribution of observing time at usually internationally operated observatories. The distribution is based on applications by individual researchers or research collaborations and decided upon in a peer review process by a panel of scientists. As astronomy rarely takes place at small observatories located at the universities anymore, extensive division of work in the generation of data and the essential role of international science policy have a growing influence on the practice of research. This evolution of astrophysics not only makes it an interesting topic for social-scientific and philosophical epistemological reflections on their own, at the same time such reflection may also yield important guidance for the challenges and future policy decisions astrophysics has to face in times of increasingly expensive research programmes and exponentially growing amounts of generated data.

In Section 2, Ian Hacking's claim of astrophysical antirealism will be taken as a starting point and a motivation to subsequently shed light on the specific methodology of astrophysics in Section 3. In Sections 4 and 5, aspects of the discussions of scientific modelling and data generation, processing and interpretation that are relevant for astrophysics will be summarized, before the article is concluded in Section 6.

\section{Astrophysical Antirealism}

Astrophysics deals with phenomena and processes that are found occurring in significantly more extreme conditions than anything that can artificially be generated in a terrestrial laboratory. The range of temperatures, pressures, spatial scales and time scales pertinent to astrophysical phenomena are larger than anything that is commonly accessible to humans by direct experience. Also, the often dominant influence of gravity sets astrophysical processes apart from terrestrial laboratory settings. While applying physical methods to cosmic phenomena, astronomers have to deal with the fact that it is impossible to interact directly with the objects of astrophysical interest. Curiously, this prohibition might not only have practical consequences for the way in which astrophysical research is done, but it may also impact the ontological status of the objects of research themselves.

In his book ``Representing and Intervening'' (Hacking, 1983) Ian Hacking tried to link scientific realism to the ability of experimenters to manipulate the object of their research. By this definition, he cannot attribute the claim of realism to astrophysical objects because direct interaction with cosmic phenomena is impossible (except to a limited degree within the Solar System). Accordingly, Hacking (1989) argues for an antirealism when it comes to cosmic phenomena. His exemplary argument cites  the ambiguities involved in interpreting the observations of theoretically predicted gravitational lensing events. If, as he claims, the universe is awash with non-detectable microlenses distorting and aberrating light as it traverses space, then all astronomical observations would be inherently untrustworthy.  Distorting influences of microlenses on the observed photon fluxed could never be disentangled from intrinsic features of the source. In this context, he revisits the Barnothys' suggestion in 1965 that quasars are not a separate class of cosmic objects but rather they are Seyfert galaxies subject to gravitational lensing. 

While an interaction with the object of research enables systematic tests of possible sources of error in experimental disciplines, this option is prohibited in astrophysics. Instead, astrophysics depends on the use of models brought in line with the observed phenomenon - a fact upon which Hacking builds his second argument against astrophysical realism. The limited nature of the modelling approach, together with the fact that cosmic phenomena can be described by a multitude of adequate but mutually contradictory models shows, according to Hacking, that astrophysics cannot make a claim to realism, if realism is understood to entail a convergence of scientific descriptions towards truth. On the basis of these considerations, in a third argument, Hacking then claims a ``methodological otherness'' for astronomy and astrophysics, as compared with other, realistic scientific disciplines. The transition to a natural science, according to Hacking, is marked by the application of experimental methods. Astronomy has not made that transition. Although astronomical technology has changed drastically since the historical beginnings of this discipline, Hacking sees the underlying methodology as having remained the same: ``Observe the heavenly bodies. Construct models of the (macro)cosmos. Try to bring observations and models into line'' (Hacking, 1989; p. 577).

There have been several replies to this rather severe view of the status of modern astrophysics, and each of his three arguments has been attacked (e.g., Shapere, 1993; Sandell, 2010). First of all, his example of microlenses does not seem to be at all representative of astrophysical research. Moreover the study of gravitational lensing was still in its infancy at that time. Today it is quite possible for astrophysicists to observationally decide on the presence (or not) of microlenses. This demonstrates the fact that often theoretical as well as technological progress can eventually resolve problems of apparent underdetermination, as is the case for Hacking's prime example given above. This point is related to the fact that observational evidence for or against a given theory does not depend on the theory alone, but it also involves and entails auxiliary hypotheses, available instruments and background assumptions (e.g. Laudan \& Leplin, 1991). If these auxiliaries are changed, the observational evidence for a theory might change as well, and subsequently give rise to differences in the observational support for previously empirically degenerate theories. Similarly, the theory itself may become further developed at a later time and allow for new predictions that can in turn be tested with new observations. For instance, in the case of gravitational microlenses, progress in our detailed theoretical understanding of lensing events in general has suggested that time-dependent brightening and fading of the emission of a microlensed background object can discriminate microlensing from properties intrinsic to the source itself. Compared with the situation Hacking describes, today we are actually able to predict observable differences between a theory that includes microlenses and one that does not. Although the argument by Laudan and Leplin (1991) does not guarantee that the problem of so-called ``contrastive underdetermination''\footnote{Contrastive underdetermination describes the idea that for each body of possible evidence there may well exist several different, but equally empirically adequate theories. Stanford (2013) contrasts this use of the term underdetermination with ``holist underdetermination'', which describes the fact that a general theory can always be held even in the light of countervailing evidence. This is because the evidence depends not only on the tested claim, but also on many other auxiliary hypotheses, which could also be equally wrong. } will always be solved sooner or later, it might explain the faith that scientists usually put in ``abductive justification'' of their science; that is, if a theory is the best possible explanation of the given empirical evidence, it might very likely be true (and if it is not, then they will find out at some later point). Accordingly, Hacking seems to have underestimated the improved understanding of astronomers and the power of their method of research. 

Hacking's second argument, referring to the ubiquitous use of models in astrophysics, has implications that reaches well beyond astronomy. Models and simulations are crucial tools in basically all fields of modern sciences. Accordingly, applying this argument would mean questioning the realist status of most of today's scientific research. Apart from that, as will be discussed in Section 4, astrophysical models are being continuously refined and improved. Moreover, new and varied observations are brought to bear in resolving ambiguities between different models. In that regard, models in astrophysics are very often critically discussed and reflected upon, and scientists are very careful with a realistic interpretation of features of these models and simulations because their limitations are usually obvious. 

Hacking's third argument concerning a fundamental difference between experimental and observational disciplines is more general than the two previous arguments, and raises two basic questions in response: (a) Does the Hacking criterion for scientific realism make sense in general? And (b) Are there really any significant epistemic differences between astrophysics and experimental sciences? The first question has been extensively discussed quite independently of its application to astrophysics (e.g. Resnik, 1994; Reiner and Pierson, 1995), and we refer the reader to that literature. Regarding the second question, a closer look at astrophysical research practice demonstrates that the distinction between experimental and observational sciences is more subtle than it seems at first sight. The crucial part is to clarify what Hacking's experimental argument explicitly means and comprises. In ``Representing and Intervening'' (1983) he gives several explanations of what it means for an entity to be real. Besides the ``strong requirement'', that direct manipulation of the entity has to be possible, a weaker version of the criterion can be found in Hacking's additional proposal that an entity can be real if its well-understood causal properties can be used to interfere with other parts of nature.

Shapere (1993) has built on this ambiguity of definition and apparent lack of conceptual clarity when he stresses that the ``use'' of cosmic phenomena in the investigation of others is indeed possible, even when the phenomena under consideration cannot be directly or interactively manipulated. Gravitational lensing, for instance, can and has been widely used for the detection of dark matter and for distance determinations. For Shapere the scientific method is not so much based on experimental interaction, but rather on the practice of extrapolating from knowledge already obtained to something novel. In this perspective, astronomy is perfectly in line with the other natural sciences. Sandell (2010) argues in the same vein when she suggests that astronomers do in fact carry out experiments, even within Hacking's meaning of manipulating or, equivalently, utilizing the causal powers of phenomena. Astronomical objects also have causal impact, which can be detected by astronomers to generate stable phenomena (e.g., measurement results of a receiver.) 	

Based on these arguments, it would appear that there is in fact no fundamental difference between astrophysics and other scientific disciplines when it comes down to scientific realism\footnote{An interesting, but unrelated, argument against a realist interpretation of astrophysical objects is based on the ambiguity and interest-dependence of astrophysical classifications. See Ruphy (2010) and Footnote 4.}. However, Hacking's article raises an interesting epistemological question. Specifically, if his argument on the possible distorting influence of microlenses is read as a variation of the strong underdetermination argument (e.g., the idea that for each body of possible evidence there might exist several different, but equally ``empirically adequate'' theories), the question remains whether experimental sciences have a larger tool box at their disposal to prevent such situations. Could it be the case that astrophysics is particularly prone to intrinsic underdetermination operating on several levels? For instance, is underdetermination found both at the level of astrophysical observations and at the level of astrophysical modelling, because of its intrinsic observational basis? In order to investigate this question, a closer look at the astrophysical method, as currently practiced, may well be in order. We do this in the next section.

\section{The Astrophysical Method}

\subsection{The Sherlock Holmes Strategy}

A significant part of astrophysical research is dedicated to the understanding of singular instances of objects (e.g., the Class 0 protostar NGC 1333-IRAS4B) or a specific process (e.g., the gaseous outflow from the active galactic nucleus in PG1211+143). The basic question in such cases is this: What do the observations tell us about the physics and chemistry at work in the observed region? Or more generally: which circumstances have led to what we observe? These scientific questions resemble corresponding questions in the ``historical sciences''. Historical research concerns itself with the explanation of existing, natural phenomena in terms of their past (and sometimes distant) causes. The fact that the causes are past means that the causal chain cannot be subject to investigation itself: the investigator finds herself in a ``Sherlock-Holmes situation''. Classical historical sciences in this sense of investigation are paleontology, archaeology and some aspects of global climate change. Historical sciences seem to differ from experimental sciences with respect to their restricted mode of evidential reasoning (e.g. Cleland, 2002). The experimental activity usually begins with a hypothesis, from which a test condition C is inferred together with a general prediction about what should happen if C is realized and the hypothesis is true. 

For instance, the hypothesis could be that astrophysicists are not interested in philosophy. The test condition C could be a lunch conversation about Kant's ``transcendental idealism'' and the prediction would be that the astrophysicist will change the topic of discussion, or leave after no more than a minute. The series of experiments then contains several experimental tests, in which C is held fixed while other experimental conditions are varied in order to exclude misleading confirmation and disconfirmation. Within the previous example, the test could be performed with a different philosophical speaker, because maybe the first speaker is intrinsically boring (independent of the topic), or the test may be rerun with a different philosophical topic than Kant, given that Kant might not be representative overall for philosophy, etc. This strategy tries to reduce the holistic underdetermination by checking different auxiliaries or confounding variables for their individual influence on the experimental prediction.

In contrast, historical sciences start from a situation that would be the experimental outcome or, in Cleland's words, they start from the traces of past causes. In the previous example, the evidence would be simply watching the astrophysicist leave the lunch table, and then the problem would be in having to reconstruct what led to this particular behaviour. Or, to take a more realistic example, if the question of the extinction of dinosaurs is being investigated, examples of such evidential traces may be a geological layer of sediment containing high levels of (extraterrestrial) iridium and the Chicxulub crater in the Gulf of Mexico, that pointed to a prehistoric asteroid impact. In astrophysics, these traces are in most cases electromagnetic photons that may be created by a wide variety of different processes. For instance, the interaction between the explosion of a supernova and a molecular cloud would lead to highly excited spectral lines that are significantly broadened by the shock interaction between the supernova remnant and the cloud. The scientific task is then to hypothesize a common, local cause for these traces (high excitation, line broadening) and thereby to unify them under one, self-consistent, causal story. In the astrophysical example, broad lines could alternatively be explained by several molecular clouds moving at different velocities along the line of sight. This possible cause of a broad spectral line would however not easily explain the existence of highly excited transitions and requires an independent causal story to explain this second trace, while a possible shock interaction unifies both.

At first sight, this ``Sherlock Holmes''-procedure seems like a very uncertain business, given the complexity of possible interactions in a potentially very long causal chain leading up to the remaining traces. However, at least there is a clear methodological direction to go in such a situation, namely to search for the so-called ``smoking guns'' (Cleland, 2002): traces that are able to discriminate between competing hypotheses, which distinguish one hypothesis as currently being the best explanation. For instance, if a shock interaction is hypothesized, other spectral lines that are predicted to arise in a shock should be observable as well. If they are not seen, alternative explanations of the already observed traces and their respective observable consequences will have to be further investigated. Doing that, the scientist has to rely on what nature has already provided, there is no way to intervene and actively create such ``smoking gun'' situations.

This fact might give rise to the intuition that historical sciences have to face a methodological disadvantage and are therefore epistemologically inferior to experimental research. However, Cleland claims that the situation is revised by a time asymmetry of nature (Lewis, 1979; Cleland, 2002) that creates an asymmetry of overdetermination. The laws of nature are directed in time: if an event has occurred, it is very difficult to make things look like nothing had happened. The situation seems analogous to criminal cases, where it is reasonable to hope that the culprit has left some traces that will make it possible to identify him. In this sense, it is difficult to fake past events, i.e. to create all expectable traces artificially without the occurrence of the actual event. While in an experimental approach, the underdetermination problem makes it difficult to find the ``true'' (responsible) causes for an observed outcome, the richness of traces in historical science cases can help to distinguish true causes from faked causes. The large number of effects that an event usually creates leads to a so called ``overdetermination of causes'' (Lewis, 1979). Even if a significant number of traces is erased, the remaining traces might be enough to determine their unifying cause. This, however, does not imply that humans will necessarily have access to these traces: ``Traces may be so small, far flung, or complicated hat no human being could ever decode them.'' (Cleland, 2002; p. 488).

Whether this alleged asymmetry really makes it easier to infer past causes (due to causal overdetermination) than to predict the future behaviour of a system (due to underdetermination of future events) as Cleland claims, seems less obvious. The large number of potential causally relevant but unidentified auxiliary factors in experimental situations may be faced with a large number of possible explanations that unify the limited set of observed traces in observational situations. Actually, in a thought experiment where an experimental setting is transformed into an observational one (by cutting the possibility of interactions with the experimental setup), it seems possible to objectively cross-link the set of possible alternative explanations with the set of possible causally relevant auxiliary factors. Accordingly, the epistemic situation in terms of underdetermination seems to be very similar for experimental and historical sciences. So far the argument was based on the ``Sherlock Holmes'' strategy as found in the historical sciences, which corresponds to the understanding of singular objects and events in astrophysics. However, the astrophysical method is obviously richer than that.

\subsection{The Cosmic Laboratory}

Although astrophysics shows some resemblances to historical sciences like paleontology or archeology, there is one fundamental difference: astrophysics does not stop at the understanding of the causal history of singular events. Instead, it ultimately tries to find general causal relations with respect to classes of processes or objects, like protostellar molecular outflows, A-type stars, spiral galaxies or galaxy clusters\footnote{Ruphy (2010) points out that it is an interesting topic on its own to study the taxonomies and classifications that are used in astrophysics in order to group entities whose diversity is scientifically investigated. Using the example of stellar taxonomies, she stresses the strong role of epistemic interest in establishing astrophysical classifications, which, e.g., usually depend on the particular observational wavelength regime, the observational resolution, and often don't have sharp boundaries when they are based, in fact, on continuous parameters. These properties of astrophysical classifications might give interesting inputs to the monism/pluralism and realism/antirealism debates, as Ruphy shows that the same structural kind-membership condition leads to several possible and interest-dependent groupings of things and that realism about stellar kinds is therefore problematic.}. This opens new methodological routes to follow: now causal relations can be investigated based on a whole class of phenomena and therefore statistical methods can be applied. In this context, astrophysicists like to refer to the ``cosmic laboratory'' (e.g., Pasachoff, 1977). The universe harbours such a variety of objects in various evolutionary states and conditions, that the experimentalist's activity of creating variations of the initial experimental condition might be already set up by the universe itself in all its naturally manifest diversity.

However, the problem remains how to preclude the influence of auxiliary factors or confounding variables on a hypothetical cause-effect relationship. How can the astrophysicist relate a certain behaviour or a certain property of a class of objects or processes rather than to the contingent environmental conditions or the specific context of the observation? That is, how can an observed correlation between observational parameters be transformed into a causal relationship?

The classic way to deal with the exploration of effects based on hypothetical causes is the ``counterfactual model'' (e.g. Shadish, Cook, and Campbell, 2002). If a certain effect is observed in a particular situation and ascribed to the existence of one particular factor, the crucial question for the claim of a causal relationship is whether the effect would have also occurred without the respective factor being in play. This method was already briefly described above as a way to deal with potentially causally relevant auxiliary factors. The perfect counterfactual inference would compare the occurring effects in two absolutely identical experimental situations that only differ with respect to the factor in question. This is however impossible. If an experiment is repeated, something else that was not mirrored by the experimenter might as well have been changed and lead to a change in the observed effect that is then incorrectly ascribed to the factor under investigation. In the experimental sciences, ``randomized controlled experiments'' yield a possible solution to this problem. Two groups of units, the so-called control and the treatment groups, are statistically similar on average but differ with respect to the factor under study. Therefore, the influence of possible confounding variables should on average cancel out for both groups and is no threat to the inferred causal relationship anymore. The challenge for astrophysics, as well as for social scientists, economists and other scientists that cannot easily perform experiments, is to find a method to deal with the influence of confounding variables that does not rely on experimental interaction with the objects of research.

Two possible methods are usually described in that context (e.g., Shadish, Cook, and Campbell, 2002; Dunning, 2012): Quasi-experiments and natural experiments. Natural experiments are the direct equivalent of randomized controlled experiments in an observational situation. The idea is that statistically similar groups of units, that only differ with respect to one factor, may sometimes be created by naturally occurring processes, which accordingly create an ``as-if random'' assignment. A classic example is the investigation of cholera transmission in nineteenth-century London (Dunning, 2012). The question whether cholera could be explained by the theory of ``bad air'' or was rather transmitted by infected water could be settled due to the particular structure of water supply in London at that time. Two different water companies served most parts of London. One obtained its water from upstream London on the Thames while the other got its water from the Thames inside of London where the water was contaminated by the sewage of London. The anesthesiologist John Snow could then show that the cholera death rate was dependent on the source of water supply. Because both groups of households were statistically equivalent on average, other reasons for the difference in the cholera death rate could be excluded. The main advantage of natural experiments is that their analysis is very simple, without necessarily calling for complicated statistical modelling. If a difference is observed between the control and the treatment group, it is most likely due to the specific factor in which both groups differ. The demanding part, however, is to decide whether the naturally created assignment is indeed ``as-if random''. This decision usually requires comprehensive qualitative information on the context of the alleged natural experiment. Also, the existence of a natural experiment is a fortunate event that might not be easy to find and that, obviously, is impossible to force.

Quasi-experiments (Campbell and Stanley, 1966), in contrast, are experiments in which a control and a treatment group are compared without random assignment being realized. Accordingly, the treatment group may differ from the control group in many ways other than the factor under study. In order to apply the counterfactual inference, all the alternative explanations that rely on other systematic differences between the groups have to be excluded or falsified. The influence of such confounding factors could in turn be investigated by dedicated experiments, but as this is usually a far too complex option to pursue. The confounding factors are often assessed by conventional quantitative methods, such as multivariate regression. Also, numerical models and simulations of the observed phenomenon can be used to explore the possible, ``covariant'', influence of these factors.  Another technique to deal with confounding factors is the application of so-called matching: the units in comparison are chosen in a way that known confounding variables can be measured and accounted for. The classical example for this method is a medical study that works with twins and can therefore rely on the far-reaching equivalence of the systems under study.

The question now becomes: Which of these methods are applied in today's astrophysical research? Does the cosmic laboratory in its vast variety offer as-if randomized assignments that allow for the targeted study of causal relationships? On first sight, a good candidate seems to be the study of evolutionary processes occurring over cosmic time-scales. Due to the finite speed of light astrophysicists are able to observe the past of the universe and thereby populations of cosmic objects in different evolutionary stages. This means that astrophysicists can literally look at the same universe at various earlier points in time. However, the objects in this earlier universe are not statistically equivalent to their present counterparts apart from their earlier age because the environmental conditions in the universe have changed in time as well. Therefore, this situation resembles the matching technique, where (hopefully) known confounding factors need to be evaluated separately, rather than it being a truly natural experiment. 

In order to evaluate the possibility of the existence of natural experiments in astrophysics more generally, one first of all needs to identify processes that may create as-if randomized assignments. The answer is not obvious at all, even though cases of so-called ``regression-discontinuity'' designs (see Dunning, 2002) might be conceivable (i.e., cases where two different groups of objects are distinguished relative to a threshold value of some variable). For instance, one could perform a study that compares molecular cloud cores that are just on the brink of gravitational collapse (expressed by the ratio of the gravitational and the stabilizing pressure forces) with those cores in the same molecular cloud that have just become gravitationally unstable. However, the existence of natural experiments in an astrophysical context seems to be hindered by the fundamental difficulty involved in obtaining enough qualitative, contextual information on the different groups of objects or processes under study. This contextual information is necessary to make a valid decision on the existence of ``as-if randomization''. In this example, the application of a threshold criterion is already rather complicated, because a decision as to which cores are just on the edge of gravitational instability is already very strongly theory-laden. Furthermore, astronomy is subject to observational constraints that introduce additional selection effects, leading to observational biases. The further the objects are away from Earth, the weaker is the received photon flux and the poorer is the spatial resolution of the object for equivalent observations (i.e., if the same telescope, same instrument, same angular resolution, and same signal-to-noise, etc. are used and obtained.) If a class of objects is observed at the same distance, it is not clear how representative those particular objects are for other regions of the universe. Also, if objects are observed in different directions, the effects of the ambient interstellar medium along the line-of-sight between the observer and the object needs to be evaluated and compensated for. Factors like this might explain why the concept ``cosmic laboratory'' is usually not associated with natural experiments and corresponding as-if randomization. The term is rather used in the following sense:  Astrophysicists can use a multitude of different phenomena, provided by the universe in various environments and evolutionary states, in order to perform\//observe quasi-experiments. However, the evaluation of these quasi-experiments requires the application of sophisticated statistical methods and the use of models and simulations, which in turn are often used as substitutes for traditional experiments.

\section{Models and Simulations in Astrophysics}

Hacking (1989) considered scientific modelling as a particularly important ingredient of astrophysical research: ``I suspect that there is no branch of natural science in which modeling is more endemic than astrophysics, nor one in which at every level modeling is more central.'' (p. 573). Whether one is willing to acknowledge such a special role for astrophysics, or not, its scientific practice does appear to be determined by modelling efforts on many different levels. The developmental time scales of many cosmic phenomena are so long that directly observing their evolution is simply not possible in a human lifetime. The accessible cosmic sample of objects in various evolutionary states is then reassembled as a self-consistent time series within a given evolutionary model, augmented and aided by simulations.

Models and simulations however always rely on idealisations, simplifications, and assumptions about the modelled object itself. Therefore the question of the reliability and validity of the results obtained by models and simulations is central. In astrophysics, unlike climate science and economics, it can be studied largely independent of political and/or public interest. As the results of astrophysical simulations cannot be tested experimentally, the only possible empirical test is a comparison with static observations. These usually result in a relatively weak measure of adequacy, because exact quantitative agreement between simulation and observation is only to be expected in exceptional cases. Alternatively, different simulations can be compared among each other, and a simulation can be tested for inner consistency and physical correctness (e.g. Sundberg, 2010).

The debate concerning models and simulations is of course an independent and vibrant activity within the philosophy of science community. Only a handful of publications, however, explicitly deal with the more philosophical or foundational aspects of modelling and simulations in astrophysics. Numerical models of astrophysical phenomena are often developed and refined over a long period of time, and can involve a large number and several generations of collaborators. A representative example of a collective and international modelling process, comprising a longstanding sequence of model-modifications, data generation and investigation of theoretical implications, was reconstructed by Grasshoff (1998). He investigated how the collective of participating researchers interacts despite their great diversity with respect to their experience, knowledge and understanding of the overall model. The split-up of the model of a phenomenon into partial sub-models, which are easier to handle, has been described by Bailer-Jones (2000). These sub-models are easier to handle, but need to be theoretically and empirically consistent if they are to be embedded at a higher level of application and generalisation. As an important aspect for the recreation of the unity of various sub-models, she identifies the visualisation of the phenomenon that supplies concrete interpretations of these sub-models. However, the often contingent choice of sub-models and the rigidity of models that grow over generations of modellers are aspects that calls for critical reflection. Using the example of cosmological simulations and modelling of the Milky Way, Ruphy (2011) stressed that many simulations are theoretically not well constrained. There are many possible paths of modelling opened by the necessary choices between different sub-models. This yields a potential or also actual plurality of models that all claim to model the same cosmic phenomenon. Typically, different models are mutually incompatible, but still empirically adequate to a similar degree. This situation creates a problem if it is possible to adjust the models to new empirical data by retroactively increasing the complexity of the new model, while keeping their previous modelling contents. This occurs in situations where no testing of the ingoing sub-models is even possible, which is often the case in astrophysics. In such a situation it becomes impossible to claim an understanding of the real world from the models just on the basis of their empirical adequacy. At best these are plausible fits to the limited input data.

That said, the current modelling practice in astrophysics appears highly heterogeneous, depending on the maturity and developmental stage of the astrophysical sub-discipline, modulated by the richness and availability of observational input. Simulations of extragalactic objects confronted with large uncertainties of ingoing theories and severe empirical constraints seem to struggle with different problems from those of the simulations of cosmic phenomena that can be observed within our Milky Way, with high spatial resolution using different information channels. 

It is important to note that models also play an important role within astronomical data acquisition and reduction. For instance, the recording of data using a single dish radio telescope requires a model of the mechanical and optical properties of the telescope mirror in different positions in order to determine the exact pointing position. For the calibration of data with respect to influences of atmospheric influences a model of the Earth's atmosphere is needed. Flux calibration presupposes models of the individual stars and planets used in the calibrating observations. As in many complex scientific disciplines, astrophysics is subject to a ``hybrid'' solution, where a clear distinction between empirical data and model-based interpretation is becoming more and more difficult. This situation might challenge elements of the modelling discussion that have occurred so far (e.g. Morrison, 2009).

Moreover, planned or newly commissioned instruments, such as the ALMA interferometer, bring about a tremendous improvement of the empirical observational basis, while at the same time the need for easy to use, standardized models is growing. Thereby, questions for the verification, validation and standardisation of simulations gain additional importance, emphasized by a common differentiation of labour between modellers and model-users.

\section{Astrophysical Data}

Astrophysics is fast becoming a science that that is confronting many large, multi-wavelength surveys and dealing with huge amounts of data. For example, the planned Square Kilometre Array will create data at a rate of many Petabytes per second. Accordingly, in addition to models and simulations, data handling is another central component of astrophysical research that is worthy of  philosophical reflection. In order to do so it too can draw on contemporary discussions from the philosophy of science.

Suppes (1962) first spoke of ``data models'', by pointing at a hierarchy of models that link raw data to theory. According to Suppes, the relationship between a data model and the underlying data is given by a detailed statistical theory of the ``goodness of the fit''. The semantic concept of models of theories is thereby extended in its application towards models of experiments and models of data. Harris (2003) has illustrated this concept of data models, using planetary observations as an example: he shows that bad observations of planetary positions are first dismissed and the remaining data points are smoothed, such that the finally plotted path of the planet would constitute a data model: on the one hand it is similar to the original data, on the other hand the plot is an idealization that is qualitatively different from the data. At the same time, the very concept of data models underwrites the inevitability of there being different ways to model data. In that sense the concept of unprocessed ``raw data'' is of questionable value surely, because ``the process of data acquisition cannot be separated from the process of data manipulation'' (Harris, 2003; p. 1512). Even for the simplest instruments it is necessary for the scientist to possess certain learned skills to properly read the data.

With their distinction between data and phenomena, Bogen and Woodward (1988) drew attention to the experimental practice within science, specifically the practice of data analysis and data selection, which had been neglected within the theory-dominated discussions of the philosophy of science. The strategies for the prevention of errors and the extraction of ``patterns'' that yield the properties of the underlying phenomena thereby constitute an epistemology of the experiment, which has become another active branch within the philosophy of science. It is interesting to apply these strategies to astrophysical data analyses. While the strategies applied are very similar to experimental physics, knowledge of the details of data generation that is needed in order to distinguish real features in the data from mere artefacts might not always be easily accessible given the ever growing differentiation of labour in observational astrophysics (e.g. Anderl, 2014).

There are several ways in which astronomical observations might actually be made (Longair, Stewart, and Williams, 1986; Jaschek, 1989; Zijlstra, Rodriguez, and Wallander, 1995). Not all of these modes are necessarily made available at any given observatory, telescope or instrument due to practicalities, cost factors and/or institutional policies. The four most common modes are: 1) Classical Observing: Here the astronomer carries out the observations herself at the telescope itself, in real time, and she is commonly assisted by a local team, consisting of telescope operators, staff scientists and instrument specialists. 2) Remote Observing. Here the local staff actually operate the telescope while the astronomer is not physically present, but she is still has the authority and responsibility to decide in real time on the target selection, integration times, filters, etc, in real time.  3) Queue Observing. In this case the instrument is operated under pre-programmed, and advanced-scheduled remote control. This is generally the case for unmanned telescopes, but it is also a mode undertaken in order to optimally schedule telescopes at many remote facilities. 4) Service Observing. In this instance, the astronomer provides the technical staff at the telescope the necessary details needed to perform a complete suite of her observations. The instructions are provided in advance of scheduled observations, they are executed without her being present or on-line, and the data are sent to her after the observations have been made.

The history seems to indicate that service observing is becoming quite popular as far as ground-based observations are concerned. However, even if the astronomer travels to the site herself, the technical complexity of the instruments involved usually requires specialized technicians to be present, who know the specifics of the instrument and how to achieve optimal performance. It is interesting that in the case of service mode observing a change of subject takes place with the concomitant possibility of information loss between the recording of raw data, and the data reduction and subsequent data analysis. In order to distinguish valid results from instrumental artefacts or blunders, it is necessary that the astronomical instruments are understood at a relatively high level and it is to be hoped that no relevant information concerning the process of data generation is undocumented or lost. Accordingly, it is a challenge for the design of archival astronomical databases to provide as much of that information on data generation and applied data manipulation as possible.

However, an epistemological analysis of data generation and data treatment has so far not thoroughly been conducted from a philosophical perspective within modern astrophysics. In contrast, the social sciences are home to a broad discussion on topics such as data-access, databases, and increasing data intensity of scientific cooperations. This discussion also refers to examples from astrophysics (e.g Collins, 1998; Wynholds et al., 2012; Sands et al., 2012). A philosophical  investigation of data generation, -selection and -reduction within astronomy appears particularly interesting, as astrophysicists like to stress their status as passive observers: they only gather information that can be received from the universe.

In fact, modern observational methods require much more manipulation and interaction than the classic picture of the astronomer looking through his or her telescope might suggest. Even before the so-called raw data is generated, many decisions have to be made, which depend on the observer's intention with respect to the usage of the data. Furthermore, the calibration of the telescope, the calibration of the data regarding the atmosphere's influence and the calibration of the receiver used are very complex processes, which rely on assumptions and models. After the user has received the data, he or she is confronted with various tasks: sorting out bad data, data calibration, searching for systematic errors and, if present, correcting for them and finally visualising the data so that a scientific interpretation is possible. Depending on the observational technique and instrumentation used, this process of data reduction can become extremely complex.

One extreme case in that respect is given by the technique of interferometry, where observations acquired by multiple telescopes are pairwise combined in order to simulate one large telescope having a resolution corresponding to the distance between the two most widely separated elements of the array of telescopes. This technique relies on the measurement of Fourier components of the source distribution on the sky, so that the intensity distribution needs to be determined by means of a Fourier transformation. At the same time, information on the underlying spatial intensity in the source plane is lost due to an incomplete sampling of the virtual surface of the simulated large telescope. The central step within the data reduction therefore cannot unambiguously reconstruct the real source distribution, but rather it can provide a plausible reconstruction, which is compatible both with the input data and with the real intensity distribution of the source on the sky. This ``data inversion'' problem might well serve as an example of true ``underdetermination'' in astrophysics.

In any case, with respect to the complexity of data generation and processing there seems to be no obvious difference between astronomy and experimental sciences. It is interesting to note that this fact is even acknowledged by Hacking (1989), although he so firmly stresses the difference between astronomy and the experimental sciences\footnote{``It is sometimes said that in astronomy we do not experiment; we can only observe. It is true that we cannot interfere very much in the distant reaches of space, but the skills employed by Penzias and Wilson [S.A.: the discoverers of the cosmic microwave background] were identical to those used by laboratory experimenters.'' (Hacking, 1983; p. 160).}.

The more complex the process of data selection and analysis becomes, the more the processing of data also relies on experience. Accordingly, instructions on astronomical data reduction are not found in textbooks and often only briefly in documentations of the respective software packages. The practice of data reduction is rather transferred in schools or workshops or directly among colleagues. This practice contains strong elements of what Polanyi (1958) described as ``tacit knowledge'', distinguished from explicit knowledge (see also Collins, 2010). The existence of tacit knowledge can become a problem if it causes a loss of information with respect to data-genesis and the context of data processing when data gets transferred. Within modern astrophysical practice, which works increasingly based on the division of labour and is directed towards a multiple use of data, the question arises how databases and corresponding data formats can be organized as to prevent a negative impact on the meaningfulness of the data.

\section{Summary and Outlook}

Modern non-cosmological astrophysics is a scientific discipline that has so far not been prominently discussed within the philosophy of science. However, the richness, complexity and extremity of its objects of research, together with its characteristic methodology as an observational science, make astrophysics an interesting field for philosophical analysis. In its attempt to understand the cosmic history of distinct singular objects and processes, it resembles historical sciences as archaeology or paleontology, while its effort to derive general claims on the behaviour of classes of objects and processes is reminiscent of social sciences, neither of which are in a position to perform experiments. All these activities, however, are backed up by far reaching modelling and simulation efforts. The question for the possible scope and inherent limits of such models appears pressing in the case of astrophysics as the possibilities of verification and validation of these models are usually limited. Astrophysics is dealing with ever growing amounts of data that are usually generated by large international observatories and finally delivered to the scientific user. The complexity of data generation, data reduction and data analysis and the reuse of data from large databases calls for a critical reflection of these practices and their epistemic premises. This article is intended to encourage philosophical interest in astrophysical research, as astrophysics may offer a wealth of novel, interesting and yet still untreated, questions and case studies that may yield new impetus to various philosophical discussions.
\vspace{0.7cm}

\noindent
{\footnotesize {\it Acknowledgements}. I would like to thank Barry Madore for his major support of my work on the philosophy of astrophysics, Paul Humphreys, who invited me to write this article, Frank Bertoldi, who encouraged my interest in philosophical reflection during my PhD program in astronomy, and last but not least Martin Harwit, who initiated my stay at the Carnegie Observatories. Furthermore, I would like to acknowledge many inspiring discussions I had during the Carnegie Cognitive Astrophysics workshops, generously funded by the Templeton foundation. }

\end{document}